\documentstyle[aps,prl,epsf,multicol]{revtex}

\begin{document}
\draft
\title{Delocalization in coupled one-dimensional chains}

\author{P. W. Brouwer$^a$, C. Mudry$^a$,
  B. D. Simons$^a$\cite{Present}, and
  A. Altland$^b$}
\address{$^a$ Lyman Laboratory of Physics, Harvard University, MA 02138, USA\\
         $^b$ Institut f\"ur Theoretische Physik, Universit\"at zu K\"oln,
Z\"ulpicher Strasse 77, 50937 K\"oln, Germany }

\date{\today}
\maketitle 
\begin{abstract}
  A weakly disordered quasi-one-dimensional tight-binding hopping
  model with $N$ rows is considered.  The probability distribution of
  the Landauer conductance is calculated exactly in the middle of the
  band, $\varepsilon=0$, and it is shown that a delocalization
  transition at this energy takes place if and only if $N$ is odd.
  This even-odd effect is explained by level repulsion of the
  transmission eigenvalues.
\end{abstract} 
\begin{multicols}{2}
\narrowtext

The existence of delocalization transitions in a disordered 
one-dimensional system is surprising, as it goes against the 
general wisdom that disordered systems in less than two dimensions are 
localized \cite{1dlocal}. 
Nevertheless a delocalization transition in one dimension 
goes back to Dyson's work on models for a glass in 1953 
\cite{Dyson1953,1drandomhopping}. 
Dyson's one-dimensional glass is 
related to a large variety 
of disordered systems:
a one-dimensional tight-binding model with nearest-neighbor random hopping 
\cite{1drandomhopping},
a two-dimensional asymmetric random bond Ising model \cite{McCoyWu}
which is equivalent to the one-dimensional random quantum Ising chain
\cite{Shankar},
one-dimensional random bond quantum XY models
\cite{Smith} and more generally random XYZ spin-1/2 Heisenberg models
\cite{Fisher},
and narrow gap semi-conductors \cite{Keldysh}.
These models are of 
current interest in view of their rich
physics: new universality classes, logarithmic scaling, the existence
of strong fluctuations calling for a distinction between average and typical 
properties. They might also be useful
laboratories to address the 
problem of
disorder induced quantum phase transitions in higher dimensions
such as the plateau transition between insulating Hall states
in the quantum Hall effect \cite{BalentsFisher,Mathur}.

The one-dimensional nearest-neighbor random hopping model 
is described by the Hamiltonian
\begin{equation} \label{eq:HamOne}
  {\cal H} = - \sum_{n} 
    \left(
    t_{n}^{\vphantom{\dagger}}c_n^{\dagger} c_{n+1}^{\vphantom{\dagger}} +
    t_{n}^{*                 }c_{n+1}^{\dagger} c_{n}^{\vphantom{\dagger}}
    \right),
\end{equation}
where the operators $c_n^{\dagger}$ and $c^{\ }_{n}$ are creation 
and annihilation operators for spinless fermions, respectively, 
and the hopping parameter 
$t_n = t + \delta t_n$ consists of a non-random part $t$ 
and a fluctuating part $\delta t_n$. 
The fundamental symmetry of the Hamiltonian (\ref{eq:HamOne}) 
that distinguishes it from one-dimensional systems with on-site disorder is 
the presence of a sublattice (or chiral) symmetry: 
particles can hop only from even to odd-numbered sites. 
The energy $\varepsilon = 0$ is special since it corresponds to a
logarithmically diverging mean density of states \cite{Dyson1953}.
Furthermore, there are several independent correlation lengths that
diverge for $\varepsilon \to 0$~\cite{Fisher} indicating that the
energy $\varepsilon=0$ represents a (disorder induced) quantum critical
point~\cite{McCoyWu,Smith,Fisher,BalentsFisher}. In particular, at
$\varepsilon = 0$ the conductance exhibits large fluctuations
superimposed on an {\it algebraically} decaying mean
value~\cite{Mathur}. By contrast, for nonzero energy the system
described by Eq.~(\ref{eq:HamOne}) is non-criticial resulting in
standard localized behaviour: A typical sample is well characterized by
$\langle \log g \rangle$, which is proportional to $L$ and has
relatively small sample- to-sample fluctuations.

A different type of delocalization in one-dimensional disordered
systems was considered recently by Hatano and 
Nelson~\cite{HatanoNelson}, who considered a chain with on-site disorder and
an imaginary vector potential. As a function of the strength of the
imaginary vector potential, the system reaches a critical point and
goes through a delocalization transition. 

The discussion so far applies to the case of {\em strictly}
one-dimensional systems. In this Letter we address the question of
whether aspects of the behaviour described above carry over to the 
multi-channel case. Surprisingly, it will turn out that the answer 
depends on the {\it parity} of the channel number $N$: For $N$ even, 
the system behaves very much like standard disordered multi-channel 
wires, i.e. in the limit $L\rightarrow \infty$ all states are localized. 
However, for $N$ odd, precisely one mode remains critical and, moreover,
exhibits much of the behaviour of the single critical mode of strictly
one-dimensional systems. For large $L$, where the contribution of all
other --- localized --- modes is neglegible, the phenomenology of the
wire is determined by the contribution of the single critical mode,
and, in this sense, remains critical.
To our knowledge, this parity effect was first noticed by Miller 
and Wang in their study of random flux and passive advection models 
\cite{WM}. However, in that work, the effect has been washed out by 
taking a two-dimensional thermodynamic limit. Keeping $N$ finite 
we here focus on a different regime, where parity has pronounced phenomenological consequences.

To probe the onset of critical behaviour as $\varepsilon \rightarrow
0$ we calculate the probability distribution of the conductance.
As will be shown below the even/odd effect manifests itself in the {\em
level repulsion} between transmission eigenvalues. We also discuss
the effect of staggering in the non-random part of the hopping
parameter $t$ (connected e.g. to a Peierls instability) and establish
a relation between delocalization transition in random hopping models
and in non-Hermitian quantum mechanics.

To be specific, we consider the Hamiltonian
\begin{eqnarray} \label{eq:HamMulti} \label{eq:HamLattice}
  {\cal H} &=& 
  - \sum_{n} \sum_{i,j=1}^{N} 
  \left(
   t_{n,ij}^{\vphantom{\dagger}}
   c_{n,j}^{\dagger}c_{n+1,i}^{\vphantom{\dagger}} +
   t_{n,ij}^{*} 
   c_{n+1,i}^{\dagger} c_{n,j}^{\vphantom{\dagger}}
   \right),
\end{eqnarray}
where the indices $i$ and $j$ label the $N$ chains. 
Weak
staggering in the hopping is introduced by setting $t_{n,ij} =
t\delta_{ij} + (-1)^n t'\delta_{ij} + \delta t_{n,ij}$, where
$t'\ll t$.  We distinguish between the cases in which time reversal
symmetry is present ($\beta=1$, $t_{n,ij}$ real) from those where it
is absent ($\beta=2$, $t_{n,ij}$ complex).
The weakly fluctuating parts of 
the hopping amplitudes $\delta t_{n,ij}$ are taken to be independent and 
Gaussian distributed, with zero mean and with variance
$  \langle \delta t_{n,ij}^{\vphantom{*}} \delta t_{n,ij}^{*} \rangle
     = \beta v^2/\gamma$,
where $\gamma = \beta N + 2 - \beta$.

Upon linearization of the spectrum in the vicinity of the Fermi energy 
$\varepsilon = 0$, the lattice model (\ref{eq:HamLattice}) can be 
approximated by a continuum model obeying the Schr\"odinger equation
\begin{mathletters}
\begin{eqnarray} \label{eq:HamCont}
  \varepsilon \psi_i(y) &=& \sum_{j=1}^{N} h_{ij}(y) \psi_j(y), 
  \ \ i=1,\ldots,N,
  \\ h_{ij} &=& 
  i v_F \delta_{ij} \sigma_1 \partial_y + v_{ij}(y) \sigma_1 + w_{ij}(y) 
  \sigma_2.
\end{eqnarray}
\end{mathletters}
Here $\psi$ is a two-component wavefunction, corresponding to even and 
odd-numbered sites in the original lattice model, $y=2na$, $a$ being the 
lattice constant, and $v_F = 2 t a$ is the Fermi velocity. 
The sublattice symmetry of the lattice model (\ref{eq:HamLattice}) 
translates to $\sigma_3 h_{ij} \sigma_3 = - h_{ij}$, 
which we refer to as chiral symmetry. 
The chiral symmetry distinguishes 
this system from one-dimensional systems 
with on-site disorder, 
which do not show a delocalization transition. 
The random potentials $v$ and $w$ are Hermitian ($v_{ij} = v_{ji}^{*}$, 
$w_{ij} = w_{ji}^{*})$ while, in the presence of time-reversal symmetry,
one has the further condition $v_{ij} = -v_{ij}^{*}$ and $w_{ij} = 
w_{ij}^{*}$. 
Apart from the symmetry constraints, the random potentials are independent 
and Gaussian distributed, with mean $\langle v_{ij}(y) \rangle = 0$ and 
$\langle w_{ij}(y) \rangle = 2 t' \delta_{ij}$, and variance 
($\bar v^2 = 2 v^2 a \beta \gamma^{-1}$)
\begin{eqnarray*}
  \langle \delta v_{ij}(y) \delta v_{ij}(y')^{*} \rangle &=& 
  \bar v^2  
  \delta(y-y') (1 - \delta_{\beta1} \delta_{ij}) 
     \\
  \langle \delta w_{ij}(y) \delta w_{ij}(y')^{*} \rangle &=& 
  \bar v^2  
  \delta(y-y') (1 + \delta_{\beta1} \delta_{ij}).
\end{eqnarray*}

In order to find the conductance at zero energy, we calculate the 
distribution of the $2N \times 2N$ transfer matrix $M$, which relates 
wavefunctions at the left and right of a disordered strip of length 
$L$ \cite{BeenakkerReview}. The eigenvalues of $M M
^{\dagger}$, which arise in inverse pairs $\exp(\pm 2 x_j)$, determine
the transmission 
eigenvalues $T_j = 1/\cosh^2 x_j$ and hence the conductance $g$ 
through the Landauer formula
\begin{equation} \label{eq:Landauer}
  g = \sum_{j=1}^{N} T_j = \sum_{j=1}^{N} {1 \over \cosh^2 x_j}.
\end{equation}
In the absence of disorder, all exponents $x_j$ are zero, 
and conduction is perfect, $g=N$. On the other hand, transmission 
is exponentially suppressed if all $x_j$'s are larger than unity. 
The $x_j$'s are related to the channel-dependent localization lengths 
$\xi_j = L/|x_j|$. The largest length $\xi$ determines 
the exponential decay of the conductance $g$ and serves as the 
localization length of the total system of coupled chains. 

To compute the distribution of $M$, we use the Fokker-Planck approach 
pioneered for disordered wires with random on-site disorder by Dorokhov 
\cite{Dorokhov} and Mello, Pereyra, and Kumar \cite{MPK}. Following the 
method of Refs.\ \onlinecite{Dorokhov,MPK}, we
first consider the case of disorder confined to a small strip 
$0 < y < \delta L$. Denoting the wave function for $y < 0$ by $\psi_j(L)$ 
and for $y > \delta L$ by $\psi_j(R)$, we find
\begin{eqnarray}
  \psi_j(R) = \sum_{k=1}^{N} M_{jk} \psi_k(L),
\end{eqnarray}
where the (random) transfer matrix $M$ of the slice reads
\begin{eqnarray}
  M = T_{y} \exp\left\{ v_F^{-1} \int_0^{\delta L} dy \left[i v(y) - 
  \sigma_3 w(y)\right] \right\}.
\end{eqnarray}
Here $T_{y}$ denotes the ordering operator for the $y$-integration.

For any given realization of the disorder, the transfer matrix has the 
following symmetry properties: 
\begin{mathletters} \label{eq:symmetry}
\begin{eqnarray}
  \sigma_3 M \sigma_3 = M &\ \ & (\mbox{chiral symmetry}), \\
  M \sigma_1 M^{\dagger} = \sigma_1 && (\mbox{flux conservation}), \\
  M^{*} = M && (\mbox{time reversal}).
\end{eqnarray}
\end{mathletters}%
Taking the symmetries (\ref{eq:symmetry}) into account, 
we find that the transfer matrix can be parameterized as
\begin{eqnarray}
  M = u \exp(x \sigma_3) v,
\end{eqnarray}
where $u$ and $v$ are the tensor product of $N\times N$ unitary matrices 
(orthogonal if $\beta=1$) with the $2 \times 2$ unit matrix
and $x$ is a diagonal $N \times N$ matrix with real diagonal 
elements $x_1,\ldots,x_N$. The numbers $x_1,\ldots,x_N$ are 
the radial coordinates of the transfer matrix
(eigenvalues of $\case{1}{2} \log M M^{\dagger}$), and the matrices 
$u$ and $v$ are the angular coordinates. In contrast to systems 
without chiral symmetry, the $x_j$ can be both 
positive and negative. 

The transfer matrix of a system of length $L$ is found by multiplication of 
the transfer matrices of the many individual slices of width $\delta L$. 
As each multiplication results in a small change of the radial coordinates 
$x_j$, they perform a ``Brownian motion'' 
\cite{Dorokhov,MPK}. Upon multiplication with the transfer matrix of a 
slice of width $\delta L$, we find that the radial coordinates $x_j$ 
change according to $x_j \to x_j + \delta x_j$, where the first two moments 
of the increment $\delta x_j$, averaged over the disorder configuration in 
the added slice, are given by
\begin{mathletters} \label{eq:dx}
\begin{eqnarray} \label{eq:dxa}
  \langle \delta x_j \rangle_{\delta L}^{\vphantom{\ }} &=& 
  {\beta \delta L \over 2 \ell \gamma} \left( - f +
  \sum_{k \neq j} \coth(x_j-x_k) \right), \\
  \langle \delta x_j \delta x_k \rangle_{\delta L}^{\vphantom{M }} &=& 
  {\delta L \over \ell \gamma}
  {\delta_{jk}}.
\end{eqnarray}
Here the mean free path $\ell$ and dimensionless staggering--disorder 
ratio $f$ read
\begin{eqnarray}
  \ell &=& v_F^2/4 v^2 a,\ \ f = \gamma t' v_F/v^2 a \beta.
\label{fdef}
\end{eqnarray}
\end{mathletters}%
The first term on the r.h.s.\ of Eq.\ (\ref{eq:dxa}) results in a 
simultaneous drift of all radial coordinates $x_j$. The second term 
describes repulsion between nearby $x_j$ in the Brownian motion process. 
The Fokker-Planck equation corresponding to Eq.\
 (\ref{eq:dx}) reads
\begin{mathletters}\label{eq:FP}
\begin{eqnarray} 
  \ell {\partial P \over \partial L} &=& {1 \over 2 \gamma} \sum_{j=1}^{N} 
{\partial \over \partial x_j} 
\left(\beta f + J {\partial \over \partial x_j} J^{-1} \right) P, \\
  J &=& \prod_{k > j} |\sinh(x_j- x_k)|^{\beta}.
\end{eqnarray}
\end{mathletters}%
The initial condition corresponding to perfect transmission at 
$L=0$ is $P(x_1,\ldots,x_N;0) = \prod_{j} \delta(x_j)$.

The Fokker-Planck equation (\ref{eq:FP}) is the central result of 
this Letter \cite{footnote}.
It contains all information on the transport properties 
of the random hopping system at zero energy. Eq.\ (\ref{eq:FP}) is 
the chiral analogue of the so-called Dorokhov-Mello-Pereyra-Kumar (DMPK) 
equation, which governs the evolution of the transmission eigenvalues of 
a disordered wire \cite{BeenakkerReview,Dorokhov,MPK}. The key difference 
between 
the two equations is the presence of ``mirror imaged'' eigenvalues $x_j$ in 
the DMPK equation, which are absent in Eq.\ (\ref{eq:FP}). 
[For wires with on-site disorder, the eigenvalues $x_j$ not only repel 
from different eigenvalues $x_k$, c.f.\ Eq.\ (\ref{eq:dxa}), but also from 
the ``mirror image'' $-x_k$; in particular, $x_j$ and $-x_j$ repel.]  

In the absence of time-reversal symmetry ($\beta=2$), the DMPK equation 
has been solved exactly by Beenakker and Rejaei \cite{BeenakkerRejaei} 
by a mapping to a problem of non-interacting fermions. Using the method 
of Ref.\ \onlinecite{BeenakkerRejaei}, we 
have been able to find an exact solution of Eq.\ (\ref{eq:FP}) for 
$\beta=2$. It reads
\begin{eqnarray}
  P &=& c(L) \prod_j \exp \left(-f x_j -x_j^2 {N \ell \over L} \right) 
\nonumber \\
  && \mbox{} \times
  \prod_{k > j} {(x_k-x_j) \sinh(x_k-x_j)}, \label{eq:exact}
\end{eqnarray}
where $c(L)$ is a normalization constant. The exact solution 
(\ref{eq:FP}) has a formal analogy to the distribution of eigenvalues 
of a random matrix: it consists of a pair interaction and a potential 
part. However, while for random matrices the eigenvalue 
interaction is quadratic, here we find a more complicated level 
repulsion. Comparing our result (\ref{eq:exact}) to the exact solution of
Beenakker and Rejaei, we note the absence of the mirror-image 
eigenvalues in the interaction and potential factors.
No exact solution of Eq.\ (\ref{eq:FP}) for $\beta=1$ could be found.

To determine whether the system is at a critical point, we investigate 
the distribution of $x_j$'s for $L \to \infty$, which can be obtained 
from Eq.\ (\ref{eq:dx}) for both $\beta=1$ and $\beta=2$. 
For $L \gg N \ell$, 
the radial coordinates $x_j$ are well separated, 
say $x_1 \ll \ldots \ll x_N$. We then find from Eq.\ (\ref{eq:dx}) 
that the ``dynamics'' of the $x_j$'s ($j=1,\ldots,N$) separate, 
and that they show small Gaussian fluctuations around equidistant 
equilibrium positions, 
\begin{eqnarray}
  \langle x_j \rangle &=& (N+1-2j-f) L \beta/2 \ell \gamma,\ \
  \mbox{var}\, x_j = L/\gamma \ell. \label{eq:crystal}
\end{eqnarray}
This is the so-called ``crystallization of transmission eigenvalues'' 
\cite{BeenakkerReview}, 
which is a signature of localization in wires with on-site disorder. 
Transmission is exponentially suppressed if {\em all} radial coordinates 
$x_j$ are larger than unity, c.f.\ Eq.\ (\ref{eq:Landauer}). 
For on-site disorder all 
$x_j$'s grow linearly with $L$ \cite{BeenakkerReview}, which inevitably 
leads to strong localization. 
[Within the framework of the DMPK equation, this results from the 
repulsion between $x_j$ and the mirror image $-x_j$.] 
The situation is different for the coupled random hopping chains, where 
we find from Eq.\ (\ref{eq:crystal}) that the radial coordinate $x_j$ 
remains (on average) close to zero, thus resulting in a delocalized state 
and a critical point, provided 
\begin{equation}
  N+1 - 2j - f = 0. \label{eq:feq}
\end{equation}
As a result, in the absence of staggering ($f=0$), 
a critical point exists {\em only if the number of chains is odd}. 
If there is no staggering, an even number of coupled random hopping 
chains show an exponential decay of the conductance. 
The conductance distribution at the critical point follows directly 
from the Landauer formula (\ref{eq:Landauer}) and the Gaussian 
distribution of the radial coordinate $x_j$. As fluctuations of $x_j$ 
around zero are large [see Eq.\ (\ref{eq:crystal})], 
the conductance at the critical point shows large sample-to-sample 
fluctuations, and the random hopping chains at the critical point 
can by no means be regarded as a ``good conductor''. 

\begin{figure}[hbt]
\centerline{\epsfxsize=1.75in\epsfbox{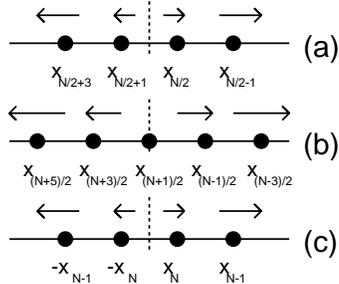}}
\smallskip
\caption{\label{fig:2} The parity effect results from level 
  repulsion between the transmission eigenvalues.  (a) For an even
  number of chains, all radial coordinates $x_j$ are repelled away
  from $0$, while (b) for an odd number of chains $x_{(N+1)/2}$
  remains close to $0$. (c) Repulsion from mirror images for a wire with
  on-site disorder results in a positive driving force for all $x_j$.}
\end{figure}

The parity effect for the presence of a critical point in the absence 
of staggering can be understood from the ``level repulsion'' of the 
variables $x_j$. In the large-$L$ limit, where $x_1 \ll \ldots \ll x_N$, 
the coordinates $x_j$ repel by constant forces, see Eq.\ (\ref{eq:dxa}). 
For an even number of channels, there is a net force on all $x_j$'s, 
driving them away from $0$ and resulting in an exponential suppression 
of the conductance (see Fig.\ \ref{fig:2}a). 
However, as is depicted in Fig.\ \ref{fig:2}b, 
if the number of channels is odd, there is no force on the middle 
exponent $x_{(N+1)/2}$. Therefore, this variable will remain close to 
zero and give rise to a diverging localization length 
$\xi = L/|x_{(N+1)/2}|$ and a critical state. For comparison, 
in the case of a wire with on-site disorder, the repulsion between 
$x_j$ and its mirror image $-x_j$ results in a nonvanishing force 
for all radial coordinates \cite{BeenakkerReview} (see Fig.\ \ref{fig:2}c). 

By fine tuning the staggering parameter $f$ (\ref{fdef}), 
which measures the ratio of 
the uniform staggering $t'$ and the random disorder strength $v$, 
an additional $[N/2]$ critical points can be reached, both for even and 
odd number of chains ($[N/2]$ is the largest integer $\le N/2$). 
According to Eq.\ (\ref{eq:crystal}), 
as the staggering parameter $f$ approaches the critical value $f=2j-N-1$, 
the localization length $\xi = \xi_j = L/|x_j|$ diverges with (critical) 
exponent $1$.

\narrowtext
 
The fact that we find exponential suppression of the conductance if 
Eq.\ (\ref{eq:feq}) is not obeyed, may be due to either the existence 
of localized states, or to a gap in the spectrum. For instance, 
in Peierls materials, staggering opens a gap in the 
excitation spectrum, which explains the exponential suppression of 
the transmission even for zero disorder. The presence of disorder 
leads to a finite density of states below the excitation gap, 
but these subgap states are localized except for critical 
realizations of the disorder strength that satisfy Eq.\ (\ref{eq:feq}). 

To close, we discuss the relation between the critical points for the 
multi-chain random hopping model studied in this Letter and the 
Hatano-Nelson delocalization transition in (one-dimensional) 
non-Hermitian quantum mechanics \cite{HatanoNelson}. The relation 
between the two systems is established through the 
``method of Hermitization'' \cite{Sommers}, in which the non-Hermitian 
problem with ``Hamiltonian'' $h$ at (complex) energy $z$ is made 
Hermitian by considering the Hamiltonian
$H_z = \sigma_1 \mbox{Re}\, (h - z) + \sigma_2 \mbox{Im}\, (h - z)$.
An eigenfunction of $h$ at eigenvalue $z$ is an eigenfunction of $H_z$
at eigenvalue $0$ and vice versa. For complex disorder in the the
non-Hermitian system, we find that the $N$-chain non-Hermitian problem
maps to $2N$ coupled chains with Hermitian quantum mechanics {\em and}
with chiral symmetry. The staggering parameter in the chiral system
plays the role of an imaginary vector potential considered by Hatano
and Nelson \cite{HatanoNelson}. Thus, comparing the non-hermitian
problem with the random
hopping chain containing an even number of rows, we deduce that, in the
absence of the imaginary vector potential, the non-Hermitian system is
localized. Since the imaginary vector potential maps to the
staggering, a series of $N$ critical points (and the corresponding
branches of delocalized states with complex energy, see Ref.\ 
\onlinecite{HatanoNelson}) can then be obtained by tuning the values
of the imaginary vector potential.

We are indebted to D. S. Fisher and B. I. Halperin for useful 
discussions. 
One of us (AA) would like particularly to acknowledge important
discussions with J. T. Chalker at an early stage of this project,
concerning both the physical background of the problem and its
formulation.
PWB acknowledges support by the NSF under grants no.\ DMR 94-16910, DMR
96-30064, and DMR 94-17047.  CM acknowledges a fellowship from the
Swiss Nationalfonds.

\end{multicols}
\end{document}